\documentclass{aa}  

\usepackage{graphicx}
\usepackage{txfonts}

\bibpunct{(}{)}{;}{a}{}{,} 

\begin{document} 

\title{Reignited star formation in dwarf galaxies quenched during reionization}

\author{E. Ledinauskas
      \inst{1,2}\thanks{E-mail: eimantas.ledinauskas@ftmc.lt}
      \and
      K. Zubovas
      \inst{1,2}
      }

\institute{Center for Physical Sciences and Technology,
          Saulėtekio al. 3, Vilnius LT-10257, Lithuania
     \and
          Vilnius University Observatory, Saulėtekio al. 9, Bldg. III, Vilnius LT-03100, Lithuania\
         }

\date{Received 13/02/2018; accepted 21/03/2018}
 
  \abstract
   {Irregular dwarf galaxies of the local group have very varied properties and star formation histories. Some of them formed the majority of their stars very late compared to the others. Extreme examples are Leo A and Aquarius which reached the peak of star formation at $z<1$ (more than 6 Gyr after the Big Bang). This fact seemingly challenges the $\Lambda$CDM cosmological framework because the dark matter halos of these galaxies on average should assemble the majority of their masses before $z \sim 2$ (<3 Gyr after the Big Bang).}
   {In this work we investigate whether the delayed star formation histories of some irregular dwarf galaxies could be explained purely by the stochasticity of their mass assembly histories coupled with the effect of cosmic reionization.}
   {We develop a semi-analytic model to follow the accretion of baryonic matter, star formation and stellar feedback in dark matter halos with present day virial masses $10^9 \, M_\odot < M_{\rm dm,0} < 10^{11} \, M_\odot$ and with different stochastic growth histories obtained using the PINOCCHIO code based on Lagrangian perturbation theory.}
   {We obtain the distributions of observable parameters and the evolution histories for these galaxies. Accretion of baryonic matter is strongly suppressed after the epoch of reionization in some models but they continue to accrete dark matter and eventually reach enough mass for accretion of baryonic matter to begin again. These ``reborn'' model galaxies show very similar delayed star formation histories to those of Leo A and Aquarius.}
   {We find that the stochasticity caused by mass assembly histories is enhanced in systems with virial masses $\sim 10^{10} \, M_\odot$ because of their sensitivity to the photoionizing intergalactic radiation field after the epoch of reionization. This results in qualitatively different star formation histories in late- and early-forming galaxies and it might explain the peculiar star formation histories of irregular dwarf galaxies such as Leo A and Aquarius.}

   \keywords{Galaxies: dwarf --
                Galaxies: formation --
                Galaxies: evolution --
                Galaxies: irregular --
                Galaxies: star formation
               }

   \maketitle
%

\section{Introduction}

The $\Lambda$CDM cosmological framework successfully explains a lot of observational facts about the cosmic microwave background \citep{Planck2015}, the large scale structure of the Universe \citep{galaxy_clustering} and galaxy formation \citep{EAGLE}. However, there are some indications that $\Lambda$CDM cosmology struggles to explain observations on smaller scales, namely on the scales of dwarf galaxies. Some of the primary disagreements between cosmological structure formation simulations and observations are the following: 1) simulations predict that galaxies like the Milky Way and M31 should have at least an order of magnitude more satellite galaxies than is observed \citep[the missing satellites problem; ][]{missing_sat_1999}; 2) a related and more troubling discrepancy is that simulations predict that galaxies like the Milky Way and Andromeda should host $\sim 6$ massive satellite galaxies which should be easily detected but are not observed in reality \citep[the too big to fail problem; ][]{too_big_to_fail}; 3) observations show that density profiles of low mass galaxies have cores in the centres instead of cusps which are predicted by cold dark matter simulations \citep[the cusp vs core problem; ][]{cuspvscore}; 4) star formation histories of some of the Local Group irregular galaxies show a delayed star formation activity which peaks at $z \lesssim 1$. In these galaxies the majority of stars form significantly later than most of the mass is assembled according to dark matter simulations \citep{leo_sfh,aquarius_sfh}. This last problem is the main motivation for this work.
	
Possibly at least some of these problems might be explained by improper comparison between observations and theory, as was shown in the context of too big to fail \citep{improp_comp_toobigtofail} and cusp vs core \citep{improp_comp_cuspvcore} issues. Nevertheless, if at least some of these inconsistencies are real, they are very important, because they might hint in favour of some corrections to the standard cosmological model \citep[e.g. warm dark matter:][]{WDM_simulations}. However, they could also be a result of various processes related to baryonic matter \citep{APOSTLE_simulations}. The latter could be especially important for dwarf galaxies as they are very sensitive to baryonic physics because of their shallow gravitational potentials. However, simulating complex physics of radiation transfer, star formation, stellar feedback and active galactic nuclei feedback from first principles requires presently unreachable numerical resources. In order to model these phenomena, simplified sub-resolution models are used which have a large number of free parameters and hence have a lower predictive power \citep{feedback_comparison}. Because of this it is still unclear whether realistic baryonic matter modelling is enough to bring dwarf galaxy formation models into agreement with observations. Therefore it is important to do more research on dwarf galaxies and check various ideas which could explain their observed peculiarities.	
		
In this work we concentrate on isolated dwarf irregular galaxies with delayed star formation histories. The most extreme examples of these are Leo A and Aquarius galaxies. In Aquarius, star formation peaks at $z \sim 0.9$ \citep{aquarius_sfh} and in Leo A it peaks at $z \sim 0.2$ \citep{leo_sfh}. In both of them, star formation decreases slowly until the present day and both of them are gas-rich \citep{Kirby2017}. On average, dwarf galaxies with present day virial masses $\lesssim 10^{10} \, M_\odot$ should assemble half of their final mass at $z \gtrsim 2$ \citep{avg_mahs_from_nbody} and their delayed star formation seems to challenge the structure formation models based on $\Lambda$CDM cosmology. However, individual mass assembly histories are stochastic and can significantly differ from the averaged one. Because the effect of heating by intergalactic radiation field and stellar feedback depends strongly on the mass of dwarf galaxies, differences in individual mass assembly histories might be amplified and result in even more different observable properties, such as stellar or gas masses and star formation histories for these individual galaxies.

In order to study the star formation histories of isolated dwarf galaxies and investigate whether the observed delayed star formation in some of them could be explained within the standard $\Lambda$CDM cosmology, we created a semi-analytic model of dwarf galaxy evolution. We generate dark matter halo merger trees by using the publicly available PINOCCHIO code based on the Lagrangian perturbation theory \citep{PINOCCHIO}. We then model galaxy evolution by supplementing these merger trees with approximate relations describing the evolution of baryonic matter. This scheme of modelling galaxy evolution in cosmological context by using precompiled merger trees is not entirely new and dates back to \citet{first_SAM_MT} and \citet{second_SAM_MT}. In the context of dwarf galaxies a similar scheme was employed already in \citet{mah_stochastika_EPS_plius_SPH}, where the authors used dark matter halo merger trees generated by an algorithm based on extended Press-Schecter theory and then used N-body and smooth particle hydrodynamics code to model the ``internal'' evolution of these halos and baryonic matter.

By using our model we investigated how the stochasticity in mass assembly histories might affect the evolution of isolated dwarf galaxies and find that, indeed, the stochasticity of mass assembly coupled with reionization leads to a wide variety of star formation histories of galaxies with dark matter halos with present-day mass $M_\mathrm{dm,0} \sim 10^9$ - $10^{10} \, M_\odot$ (the characteristic transitional mass below which the accretion of the baryonic matter is strongly suppressed and above which it is suppressed weakly). This results in qualitatively differing star formation histories in the models of late- and early-forming dwarf galaxies. We find that in some late-forming model galaxies star formation is temporarily suppressed by cosmic reionization but, because they still can accrete dark matter as it is not affected by radiation, eventually they become massive enough to again start accreting baryonic matter and thus begin to form stars once more. Such ``reborn'' galaxies have been already numerically obtained and analysed in \citet{dopey_and_grumpy, reborn_dwarfs_1, reborn_dwarfs_2}. We show that some of the reborn galaxies in our models have very similar star formation histories to Leo A and Aquarius and therefore we conclude that stochasticity of mass assembly might explain their peculiar observed properties.

The model is described in section 2. In section 3, we present the main results. Implications and credibility of the results are discussed in section 4. The main conclusions are summarized in section 5. In this work we assume a standard $\Lambda$CDM cosmology with cosmological parameters $H_0 = 67.74 \, \mathrm{km} \, \mathrm{s}^{-1} \mathrm{Mpc}^{-1}$, $\Omega_\mathrm{b} = 0.05$, $\Omega_\mathrm{m} = 0.31$ and $\Omega_\Lambda = 0.69$ from \citet{Planck2015}.

\section{Description of the model}
	Our model is very similar to other semi-analytic galaxy evolution models \citep[e.g.,][]{galform,sage, galICS}. Main difference is that we focus on isolated dwarf galaxies and so evade the necessity to model AGN feedback and interactions with more massive galaxies. The main components of each model galaxy are a dark matter halo, a gas disk and a stellar disk. We model their evolution by using dark matter halo merger trees and simplified analytical relations which dictate mass changes due to accretion, mergers, star formation and feedback across the three components.

	\subsection{Dark matter halo merger trees}
		
	To generate dark matter halo merger trees we use the publicly accessible PINOCCHIO code \citep{PINOCCHIO}. It is an approximate tool used for dark matter halo catalogs with spatial information at various redshifts based on the Lagrangian perturbation theory. Merger trees obtained from N-body simulation would be more accurate, but the possible inaccuracies of this simplified approach should not be important compared to inaccuracies caused by approximations that we have to make when modelling complex baryonic physics. Therefore we choose this method because it runs orders of magnitude faster than direct N-body simulations so the computing power required to obtain the merger trees of sufficient resolution to model low-mass galaxies can be obtained even on a desktop computer. In this work we use a run with box size $d \approx 28 \, \mathrm{Mpc}$, sampled with $800^3$ particles and with periodic boundary conditions. If we assume that the minimal halo comprises of 10 particles (standard value for PINOCCHIO) then its mass is $M_\mathrm{min} \approx 2 \times 10^7 \, M_\odot$ which is enough to resolve merger trees of halos used in this work ($M_\mathrm{dm,0} \geq 10^9 \, M_\odot$). 
		
\begin{figure}
			 	\centering
			 	\includegraphics[width=9.2cm]{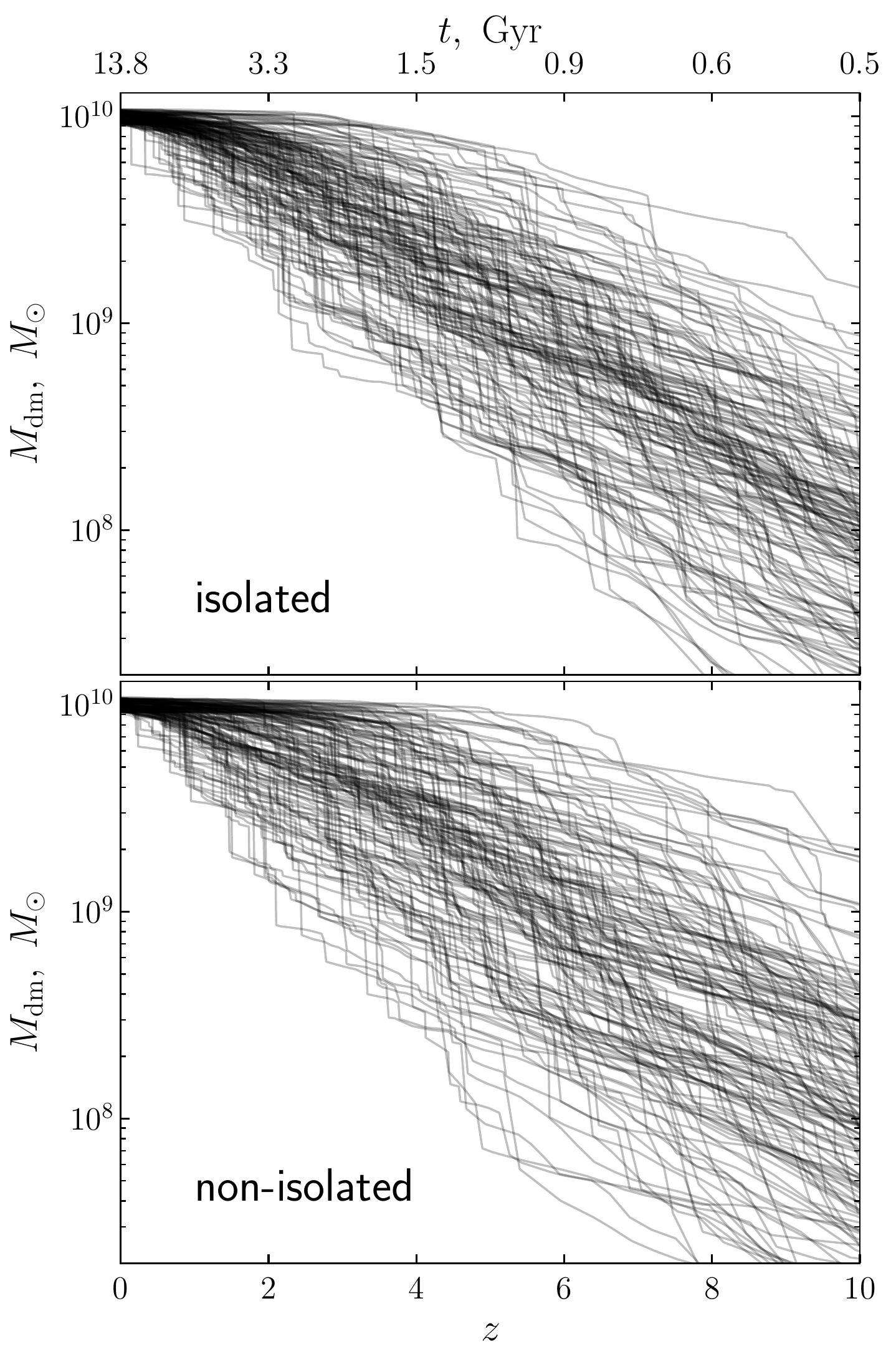}
			 	\caption{Example dark matter mass assembly histories obtained with PINOCCHIO for isolated (top) and non-isolated (bottom) halos which all have $\approx 10^{10} \, M_\odot$ (within 10\%) mass at $z=0$.}
			 	\label{growth_histories}
		 \end{figure}

	In this work we concentrate on isolated dwarf galaxies, so out of all dark matter halos we only analyse those which are isolated throughout their whole evolution. We define a halo as isolated if it is further than $2 r_\mathrm{vir}$ from all more massive halos, where $r_\mathrm{vir}$ is the virial radius of the more massive halo. We use a common $r_\mathrm{vir}$ definition:
	\begin{equation}
		r_\mathrm{vir} = \bigg( \frac{3 M_\mathrm{dm}}{4 \pi \rho_\mathrm{crit} \Delta_\mathrm{vir}} \bigg)^{1/3} \,,
	\end{equation}
	where $M_\mathrm{dm}$ is dark matter halo mass, $\rho_\mathrm{crit}$ is the critical density of the Universe and $\Delta_\mathrm{vir}$ is the overdensity of the collapsed and virialized spherical top-hat density fluctuation. We use a fitting formula from \citet{Delta_vir} to get the value of $\Delta_\mathrm{vir}$, which changes from $\Delta_\mathrm{vir} \approx 178$ at $z > 5$ to $\Delta_\mathrm{vir} \approx 102$ at $z = 0$. Figure \ref{growth_histories} shows examples of different mass assembly histories of isolated and non-isolated halos which all have a present-day mass $M_\mathrm{dm,0}\approx 10^{10} \, M_\odot$ to within 10\%. As can be seen from the figure \ref{growth_histories}, isolated halos accumulate their mass on average slightly later than non-isolated ones. The median age of the Universe at which a $M_\mathrm{dm,0} \approx 10^{10} \, M_\odot$ halo accumulates 90\% of its final mass is $t_{90} \approx 6.2 \, \mathrm{Gyr}$ for isolated halos and $t_{90} \approx 5 \, \mathrm{Gyr}$ for non-isolated ones.  

	\subsection{Dark matter halos}
	
		We assume that densities of dark matter halos follow the spherically symmetric NFW profile \citep{nfw}:
		\begin{equation}
			\rho_\mathrm{dm}(r) = \frac{\rho_0}{r / r_\mathrm{s} (1 + r / r_\mathrm{s})^2} \,,
		\end{equation}
		
		\noindent where $r_s$ is the scale radius. With this profile the total mass of the halo diverges as $\ln(r/r_\mathrm{s})$ so we truncate it at the virial radius. The scale radius and virial radius are related via the concentration parameter $c \equiv r_\mathrm{vir} / r_\mathrm{s} $. The concentration parameter of a particular halo depends on its mass assembly history. To calculate the concentration parameter at a given time $t$ we use the model of \citet{dm_halo_concentration} which relates $c$ to the time, $t_{0.04}$, at which the halo assembled 4\% of the mass it has at time $t$:
		\begin{equation}
			c(t) = 4 \bigg[ 1 + \bigg( \frac{t}{3.75 t_{0.04}} \bigg)^{8.4} \bigg]^{1/8} \,.
		\end{equation} 
		According to this equation, $c$ approaches 4 when a halo is rapidly growing due to dynamical heating and it increases when the halo is growing slowly due to relaxation.
		
		To characterize the gravitational potential of the dark matter halo we use its maximum circular velocity, which for an NFW profile is:
		\begin{equation}
			v_\mathrm{max} \approx 0.465 r_\mathrm{s} \sqrt{4 \pi G \rho_0} \,.
		\end{equation}

	\subsection{Baryonic matter accretion} \label{baryon_accretion}
		The mass of the dark matter halo grows in two modes: smooth accretion from intergalactic medium and mergers with other halos. We assume that before the reionization epoch the smooth baryonic matter accretion rate is always proportional to the smooth dark matter accretion rate:
		\begin{equation}
			\dot{M}_\mathrm{g,acc} = \langle f_\mathrm{b} \rangle \dot{M}_\mathrm{dm,acc} \,,
		\end{equation}
		\noindent where $\langle f_\mathrm{b} \rangle = \Omega_\mathrm{b} / \Omega_\mathrm{dm} \approx 0.19$ is the ratio between the average baryonic and dark matter densities in the Universe. During the epoch of reionization, the average temperature of intergalactic gas rapidly increases to $\langle T_\mathrm{IGM} \rangle\approx 2 \times 10^4 \,\, \mathrm{K}$ \citep{reionization_temp}. Therefore low mass halos are prevented from accreting baryonic matter \citep{gnedin2000} and lowest mass halos even lose their gas due to photoevaporation. In order to account for these effects, a simplified algorithm can be used as shown in \citet{okamoto2008}. We use basically the same algorithm in this work. If the halo virial temperature $T_\mathrm{vir} < T_\mathrm{eq}(\Delta_\mathrm{vir} / 3)$, the halo does not accrete any gas. Here $T_\mathrm{eq}(\Delta_\mathrm{vir} / 3)$ is the temperature of intergalactic gas at overdensity $\Delta_\mathrm{vir} / 3$. Virial temperature in this work is defined as:
		\begin{equation}
			 T_\mathrm{vir} = \mu m_\mathrm{p} v_\mathrm{max}^2 / (2 k_\mathrm{B}) \,,
		\end{equation}
		 where $m_{\rm p}$ is the mass of a proton and $\mu \approx 0.63$ is the average mass of a gas particle in units of $m_\mathrm{p}$. Note that we use $v_\mathrm{max}$ instead of more commonly used virial velocity $v_\mathrm{vir} = \sqrt{G M_\mathrm{dm} / r_\mathrm{vir}}$. We do that because when $r_\mathrm{vir}$ is defined by (1) it increases in time even for a fixed halo mass because of decreasing $\rho_\mathrm{crit}$ and this leads to an unphysical decrease in $v_\mathrm{vir}$. $v_\mathrm{max}$ takes the halo concentration into account and so is less sensitive to this effect. For the lowest mass halos, if $T_\mathrm{vir} < T_\mathrm{eq}(\Delta_\mathrm{evp})$, where $\Delta_\mathrm{evp} = 10^6$, the gas of the halo is evaporated exponentially at a rate:
		\begin{equation}
			 \dot{M}_\mathrm{evp} = \frac{M_\mathrm{g}}{r_\mathrm{vir} / c_\mathrm{s} (\Delta_\mathrm{evp})} \,,
	    \end{equation}
	     where $M_\mathrm{g}$ is the mass of the gas disk and $c_\mathrm{s}(\Delta_\mathrm{evp})$ is the sound speed at temperature $T_\mathrm{eq}(\Delta_\mathrm{evp})$. We adopt $T_\mathrm{eq}(\Delta_\mathrm{vir} / 3)$ and $T_\mathrm{eq}(\Delta_\mathrm{evp})$ dependencies also from \citet{okamoto2008}, according to which $T_\mathrm{eq}(\Delta_\mathrm{vir} / 3)$ changes from $\approx 10^4$ K at $z = 9$ to $\approx 4 \times 10^4$ K at $z = 0$, while $T_\mathrm{eq}(\Delta_\mathrm{evp}) \approx 10^4$ K always.

		We treat galaxy mergers very approximately and assume that the larger halo just accretes all the gas and stars from the smaller one in one dynamical time. In reality mergers of galaxies with comparable mass may induce dynamical disc perturbations which might affect its structure and may lead to starbursts. However inclusion of these highly complex processes in our phenomenological model would require an introduction of free parameters which we try to avoid. So we choose to not treat major mergers in any special way. However, in our model the star formation rate during mergers still increases because of the increase of gas mass and therefore gas density (if the accreted halo had any gas left beforehand). To study how this approximation could affect the results, we calculated models in which mergers with 30\% or higher mass ratios induce extreme starbursts during which a fraction $f_* = 1/(1 + f_{\rm ej})$ of all the gas mass is turned to stars and the remaining fraction is ejected, where $f_{\rm ej} = m_{\rm ej} / m_*$ is the ratio between ejected mass by feedback of a formed stellar population and the mass of that population (see section 2.5 below). We discuss the results of this experiment in section 4. 
        
        Numerical simulations and analytic arguments show that for low mass galaxies gas accretion should be predominantly in cold mode without formation of a virially shocked hot halo \citep{bimodal_gas_accretion2, bimodal_gas_accretion, cool_infall}. Therefore we assume that the accreted baryonic mass falls straight to the central rotationally supported gas disk in a dynamical time of the dark matter halo. In order to calculate the size and density of the gas disk we use an empirical relation by \citet{Kravtsov2013}:
		\begin{equation}
			R_{*,1/2} = 0.015 r_\mathrm{vir} \,,
		\end{equation}
		where $R_{*,1/2}$ is the half-mass radius of the stellar disk. We assume that the surface density profiles of both gas and stellar disks are exponential:
		\begin{equation}
		\Sigma(R) = \Sigma_0 \exp(-R/R_\mathrm{s}) \,,
		\end{equation}
		where the scale radius is given by $R_\mathrm{s} = R_{1/2} / 1.678$. Observations show that usually scale lengths of gas disks are larger than those of stellar disks (e.g. see table 2 in \citet{m_s_m_dm_relation}). In absence of any good theoretically motivated relations, we just assume that scale length of gas disk $R_\mathrm{s,g} = 2R_{s,*}$. This is close to average value in table 2 of \citet{m_s_m_dm_relation}. We also experimented with different values of this parameter and discuss the results in section 4.

	\subsection{Star formation} \label{starform}

		To calculate the star formation rate (SFR) in the gas disk we use a physically motivated analytic model by \citet{KrumholzSFLaw} which predicts star formation law in molecule-poor galaxies and matches observations well. In this model the star formation rate surface density is calculated by:
		\begin{equation} 
		\dot{\Sigma}_* = f_{\mathrm{H}_2} \epsilon_\mathrm{ff} \Sigma_\mathrm{g} / t_\mathrm{ff} \,,
		\end{equation}
		where $\epsilon_\mathrm{ff} \approx 0.01$, $t_\mathrm{ff}$ is the free-fall time of the molecular gas and $f_{\mathrm{H}_2}$ is the mass fraction of gas in molecular form.  $f_{\mathrm{H}_2}$ in this model depends on surface density and metallicity of the gas and also on volume density of stars and dark matter. This model has a free parameter $f_\mathrm{c}$, called ``clumping factor'', which is defined as the ratio of surface densities characteristic of atomic-molecular complexes to the surface density averaged over the resolved scale. As suggested in \citet{KrumholzSFLaw}, we use $f_\mathrm{c}=5$ because in the presented model we only follow the average structure of the gas disk. In this star formation rate model it is also assumed that molecules form only on dust grains, so it cannot be used to calculate SFR in gas with metallicity $Z = 0$. To get around this we follow \citet{krumholz_bathtub} and assume that population III stars enrich the surrounding gas to $Z_\mathrm{prim} = 2 \times 10^{-5}$, as calculated in \citet{Z_prim}, and use this as a starting metallicity for primordial gas. We leave all the other parameters at their default values as described in \citet{KrumholzSFLaw}.
		
	\subsection{Stellar feedback}
		Newly formed stars inject momentum and energy into the surrounding gas via stellar winds, radiation and supernova explosions. This feedback regulates star formation efficiency on molecular cloud scales and can create massive gas outflows on galactic scales, which negatively affect the star formation rate. The latter effect is especially important for dwarf galaxies which have low escape velocities and it is necessary to include feedback into galaxy formation and evolution models in order to match the observed galaxy luminosity function \citep{semi_analytic_review}. However, stellar feedback is a very complex subject and it is far from understood. Different implementations of it give different results \citep{feedback_comparison}.    
	
		In this work we explicitly model only feedback coming from supernova explosions. Small scale effects of stellar feedback are included implicitly in the empirical star formation efficiency parameter $\epsilon_\mathrm{ff}$ (see section \ref{starform}). We model supernova feedback very approximately by using momentum conservation and assuming that every supernova ejects a gas mass 
		\begin{equation}
			m_\mathrm{ej} = p_\mathrm{SN} / v_\mathrm{max}
		\end{equation}
        from the galaxy. Here $p_\mathrm{SN}$ is the momentum generated by a supernova (including not only the momentum generated by initial stellar matter ejection but also by the subsequent adiabatic expansion during the Sedov phase until the energy losses by radiation become non-negligible). We set $p_\mathrm{SN} = 2 \times 10^{43} \, \mathrm{g \, cm / s}$ according to simulations of supernovae in a turbulent interstellar medium \citep[e.g.,][]{p_SN1,p_SN2}. We use momentum instead of energy because a large and uncertain amount of the injected kinetic energy is radiated away while the momentum remains conserved after the Sedov phase. So using energy conservation in a similar fashion would be even more uncertain. Of course eq. (11) gives approximately the maximum possible ejected mass and in reality it is almost certainly lower. However, we do not include any other feedback mechanisms on large scales, so we expect that this overly strong supernova feedback somewhat compensates for that. 
        
        The supernova rate is calculated as follows. We assume a \citet{Kroupa2001} initial mass function for the stellar population and further assume that only stars with $8 \, M_\odot < M_\mathrm{star} < 25 \, M_\odot$ explode as core collapse supernovae, as more massive stars with low metallicity typical to dwarf galaxies should experience collapse into black holes without any strong explosion \citep{massive_stars_end}. This gives about 1 supernova per $110 \, M_\odot$ of stars formed. We further add Type Ia supernovae, which have a rate of $\sim 1$ per $1000 \, M_\odot$ of stars formed \citep{SNIA_rate}, to arrive at a supernovae rate of 1 per $100 \, M_\odot$ of stars formed in the disc.

	\subsection{Chemical evolution}

	In order to model the enrichment of the interstellar medium by heavy elements and gas recycling we use data from the publicly available simple stellar population synthesis code SYGMA \citep{SYGMA}. Heavy elements are injected directly into the gas disk and then are ejected or converted to stars in proportion to the involved gas mass. We assume that the ejected heavy elements are rapidly mixed in the interstellar medium and always treat the gas disk as chemically homogeneous, having a uniform average metallicity. We also make the common approximation that heavy elements are ejected in the same time step as their parent stellar population is formed (instantaneous recycling). We use time steps $\Delta t \approx 50 \, \mathrm{Myr}$ which are longer than the life time of massive stars which explode as supernovae.
        
\section{Results}
	We calculated models for 6 samples of dark matter halos, each consisting of about $300$ isolated halos with present-day masses $M_\mathrm{dm,0} \approx 10^9 \, M_\odot$, $2.5 \times 10^9 \, M_\odot$, $5 \times 10^9 \, M_\odot$, $10^{10} \, M_\odot$, $2 \times 10^{10} \, M_\odot$, $5 \times 10^{10} \, M_\odot$ and $2 \times 10^{11} \, M_\odot$ (dispersions of these masses can be seen in fig. \ref{compare_to_obs}, panel a. We choose this mass interval because we expect that both Leo A and Aquarius fall into it based on abundance matching results \citep{Moster_abund_matching, m_s_m_dm_relation}. This interval is also interesting because in it the transition from strongly affected by reionization to weakly affected happens as halos in it have virial temperatures comparable to $T_{\rm eq}(\Delta_\mathrm{vir} / 3)$ (see section 2.3) throughout their history.

	\begin{figure*}
		\resizebox{\hsize}{!}
		{\includegraphics[]{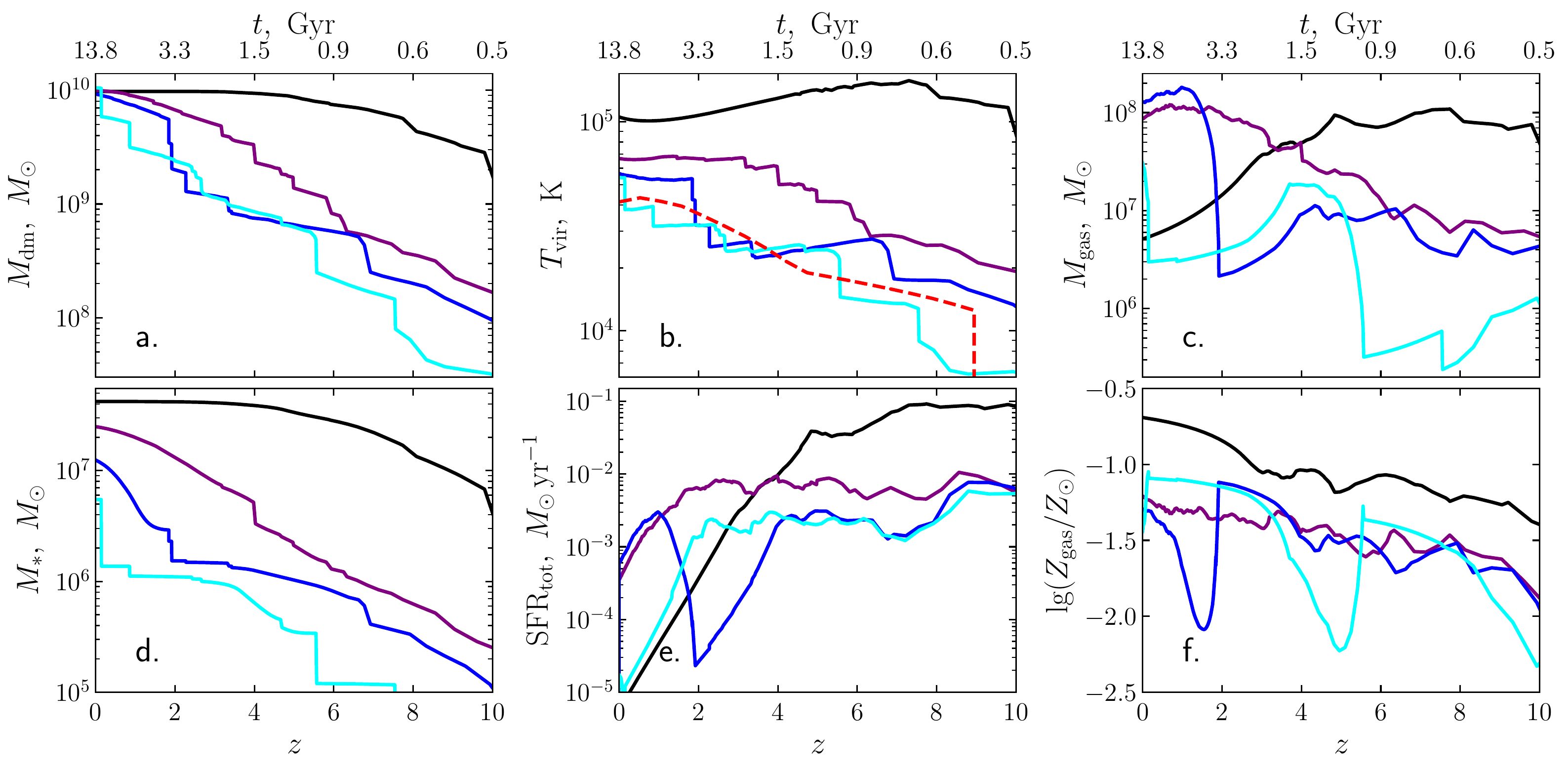}}
		\caption{Dependence on redshift of the dark matter halo virial mass (a), the virial temperature (b), the gas mass (c), the stellar mass (d), the star formation rate (e) and the gas metallicity (f) for 4 halos with different mass assembly histories and nearly the same final halo mass (within 10\%). In the panel b the red dashed line shows $T_\mathrm{eq}(\Delta_\mathrm{vir}/3)$, the virial temperature below which smooth accretion of baryons is shut off.}
		\label{pic_evolutions}
	\end{figure*}
	
	\begin{figure*}
		\resizebox{\hsize}{!}
		{\includegraphics[]{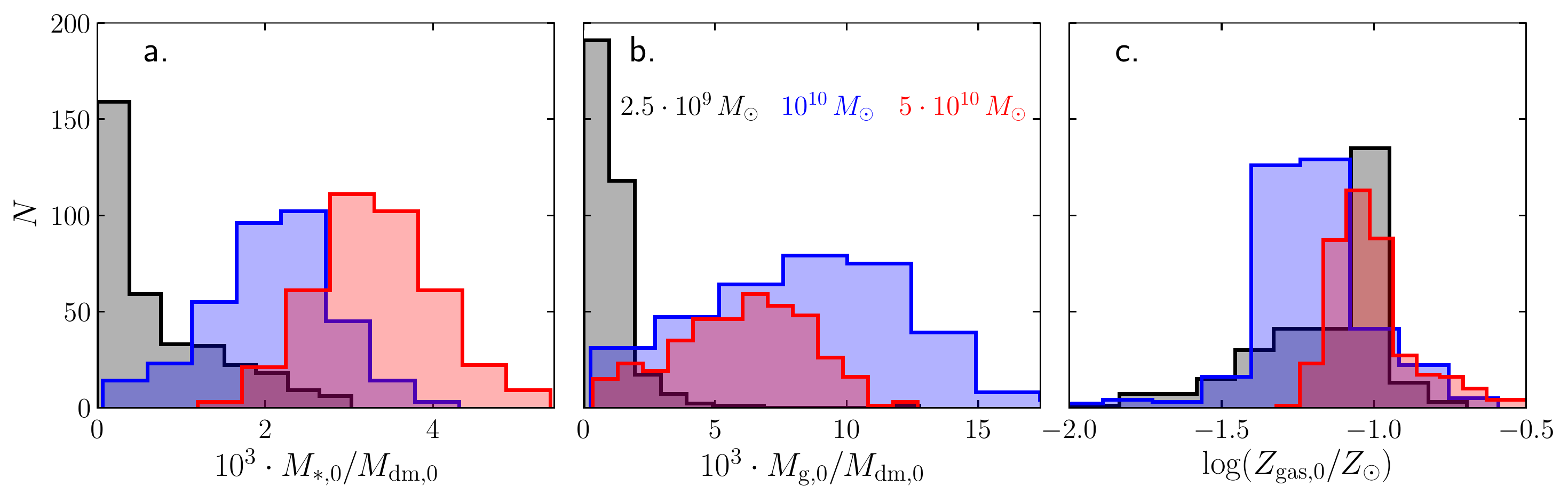}}
		\caption{Histograms of stellar mass (a), gas mass (b) and gas metallicity (c) at $z=0$ for models with present-day dark matter halo masses $2.5 \times 10^9 \, M_\odot$ (black), $10^{10} \, M_\odot$ (blue) and $5 \times 10^{10} \, M_\odot$ (red).}
		\label{histograms}
	\end{figure*}

	\subsection{Evolution of parameters in time}
	The variety of galaxy evolution histories caused by the stochasticity of mass assembly can be seen in fig. \ref{pic_evolutions}. This figure shows the dependence on time of the virial dark matter halo mass (a), the virial temperature (b), the gas mass (c), the stellar mass (d), the total star formation rate (summed across all the branches of the merger tree) (e) and the average gas metallicity (f) for four $M_\mathrm{dm,0} \approx 10^{10} \, M_\odot$ halos with very different mass assembly histories. We display the mass-weighted average gas metallicity throughout the whole disk $Z_{\rm gas} = m_{\rm met} / m_{\rm gas}$, normalized to Solar metallicity from \citet{asplund_solar_abund}. The dashed red line in panel b shows the virial temperature below which smooth baryonic matter accretion is turned off $T_\mathrm{eq}(\Delta_\mathrm{vir}/3)$ (see section 2.3). These four mass assembly histories are chosen because they represent the whole variety quite well. Note that except for panel e the plots in fig \ref{pic_evolutions} only show the parameters of the main halos in the merger trees. 
	
	The black line shows a halo which becomes massive very early and then slowly grows almost only by smooth accretion. The purple line shows a less extreme example of a halo with mass assembly history close to the median one. Both of these halos have acquired enough mass by the start of reionization era to not be affected strongly by reionization (the only effect is that less baryonic matter is accreted during mergers with other smaller halos which might be affected). Both of them have peak star formation rates at $z > 2$ and have formed 90\% of their stellar mass already at $z \approx 5 $ and $z \approx 1$ respectively.

    The blue line shows a halo which acquires its mass later and therefore experiences a period lasting $\sim 2 \, \mathrm{Gyr}$ during and after reionization when there is no smooth accretion of baryonic matter (we subsequently call this the "quenching period"). Because of quenching the star formation rate in this galaxy quickly declines after $z \approx 4$. However, after $z \approx 2$ it starts to increase again and reaches maximum at $z \approx 1$ because only at $z \approx 2$ the galaxy acquires enough dark matter mass to start accreting baryonic matter again. This results in a galaxy with qualitatively different star formation history and dominated by young stellar populations as $\approx 90\%$ of stellar mass is formed after $z = 1$.
    
	The cyan line shows an extreme example of a halo which acquires most of its mass at $z < 1$ and therefore experiences a longer quenching period lasting $\sim 10 \, \mathrm{Gyr}$. As smaller halos that merge with it are also quenched, this halo acquires very small amounts of gas compared to accreted dark matter and so star formation rate remains very low up until $z=0$. This results in a galaxy dominated by old stellar populations similarly to the first two example galaxies.
	
	In fig. \ref{pic_evolutions}, c we see that the final gas mass depends non-monotonically on halo formation time. Extremely early- and late-forming halos have present-day gas mass $M_\mathrm{gas,0} \approx 5 \times 10^6 \, M_\odot$ and $M_\mathrm{gas,0} \approx 3 \times 10^7 \, M_\odot$ respectively (large jump in gas mass of late-forming halo happens because it merged with another gas-rich halo right before $z=0$ and after that experienced rapid accretion), while those with less extreme growth histories have about an order of magnitude higher gas masses. Final stellar mass on the other hand decreases monotonically with formation time as can be seen in fig. \ref{pic_evolutions}, d. The most early forming galaxy has $M_{*,0} \approx 4 \times 10^7 \, M_\odot$ while the most late forming galaxy has $M_{*,0} \approx 5.5 \times 10^6 \, M_\odot$. By combining these two gas and stellar mass dependencies on formation time it is easy to understand the dependence of gas metallicity, which can be seen in fig. \ref{pic_evolutions}, f. The earliest-forming galaxy has the highest metallicity because it has the largest final stellar mass but low final gas mass so this results in a higher amount of ejected metals contained in a lower amount of gas. This is, of course, only a rough explanation because the model is not a closed box: gas can leave the halo and fall into it. 

    Out of all the models with $M_\mathrm{dm,0} \approx 5 \times 10^{9} \, M_\odot$, $10^{10} \, M_\odot$ and $2 \times 10^{10}\, M_\odot$ there are respectively about 70\%, 33\% and 21\% of halos in which baryonic accretion is shut off at least for one time step (50 Myr). Additionally, there are respectively 68\%, 17\% and 4\% of halos which are quenched for at least 500 Myr and 52\%, 2\% and 0.3\% of halos which are quenched for at least 5 Gyr. This rapidly changing fraction with increasing mass shows that $M_\mathrm{dm,0} \sim 10^{10} \, M_\odot$ is indeed the characteristic transitional mass between halos that are weakly and strongly affected by reionization. Note that these fractions are for the main halos in the merger trees and do not include less massive halos which are affected more. At this mass range the stochasticity of the mass assembly is amplified because early-forming halos are weakly affected by reionization while late-forming halos are strongly affected. And so, as can be seen in fig \ref{pic_evolutions}, e, this might result in galaxies which have qualitatively different star formation histories even though their dynamical masses at $z = 0$ are the same.
	
	\subsection{Distributions of final parameters}
	The histograms of the final stellar mass (a), the gas mass (b) and the gas metallicity (c) at $z=0$ for samples with $M_\mathrm{dm,0} \approx 2.5 \times 10^9 \, M_\odot$, $10^{10} \, M_\odot$ and $5 \times 10^{10} \, M_\odot$ are shown in fig. \ref{histograms} (the masses are normalized to their corresponding dark matter halo mass). As expected, stellar mass on average grows super-linearly with the halo mass, so the ratio of stellar mass to halo mass grows with halo mass. Gas mass increases super-linearly only up to $M_\mathrm{dm} \sim 10^{10} \, M_\odot$ and then increases sub-linearly. Super-linear growth happens because up to $M_\mathrm{dm,0} \sim 10^{10} \, M_\odot$ the fraction of quenched halos decreases and so they can accrete on average super-linearly larger amounts of gas. This effect becomes unimportant for more massive halos than $M_\mathrm{dm,0} \sim 10^{10} \, M_\odot$. Sub-linear mode happens because more massive halos have higher star formation densities and so they consume and eject higher fractions of their total gas mass. Metallicity between these three masses does not change much. The lower average metallicity of $M_\mathrm{dm,0} \approx 10^{10} \, M_\odot$ models is a result of higher gas-to-stellar mass ratio in them.
	
\begin{figure}
		 	\centering
		 	\includegraphics[width=8cm]{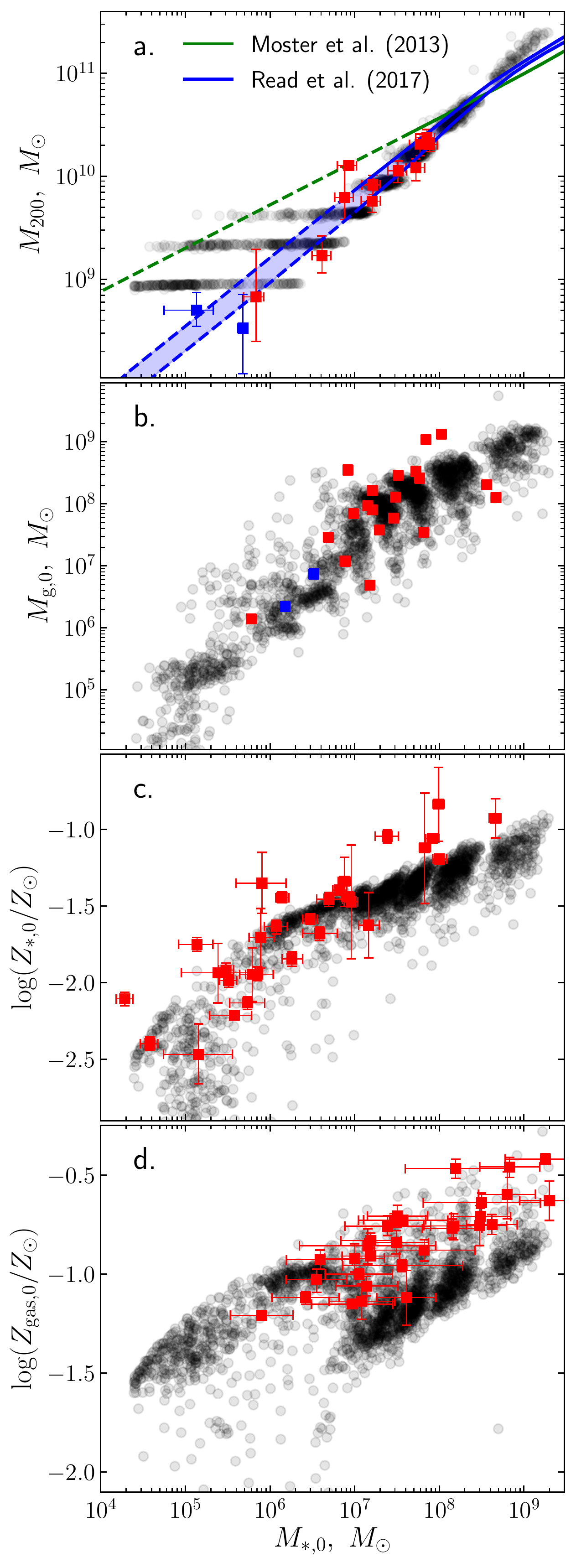}
		 	\caption{Relations between present-day stellar mass and dark matter halo virial mass (a), gas mass (b), average stellar metallicity (c) and average gas metallicity (d). Black points represent models and red points represent observations (a: \citet{m_s_m_dm_relation}, b: \citet{cuspvscore, Kirby2017}, c: \citet{m_s_Z_s_relation}, d: \citet{m_s_Z_g_relation}). Blue points in panel a represent non-isolated galaxies Carina and Leo T, while in panel b they represent Leo A and Aquarius galaxies, which are the focus of this work. Green and blue lines show $M_{*,0}$ - $M_{\rm dm,0}$ relations from abundance matching \citep{Moster_abund_matching, m_s_m_dm_relation}, dashed lines denote extrapolation.}
		 	\label{compare_to_obs}
	 \end{figure}

	The relations between stellar mass and dark matter halo mass (a), gas mass (b), gas metallicity (c) and stellar metallicity (defined analogously to $Z_{\rm gas}$) (d) for all the models together with corresponding data from observations \citep{m_s_m_dm_relation, cuspvscore, Kirby2017, m_s_Z_s_relation, m_s_Z_g_relation} are shown in fig \ref{compare_to_obs}. For easier comparison we converted dark matter halo virial masses from our model to $M_{200}$, which is defined similarly as in this work, but $\Delta=200$ is used instead of $\Delta_{\rm vir}$ (this definition is used in the cited works). Clustering of model data in these plots arises because the total sample is not continuous in $M_\mathrm{dm}$ but consists of 7 sub-samples with specific $M_\mathrm{dm}$. 
	
	In the panel a. of fig. \ref{compare_to_obs} a bimodal distribution of stellar masses can be seen for galaxies with $M_\mathrm{dm,0} \approx 2.5 \times 10^9 \, M_\odot$. This bimodality already starts to develop in halos with $M_\mathrm{dm,0} \approx 10^{10} \, M_\odot$ and starts to vanish in halos with $M_\mathrm{dm,0} \approx 10^{9} \, M_\odot$. This happens because this mass interval corresponds to galaxies in transition from weakly affected to strongly affected by reionization. In this interval, as can be seen in fig. \ref{pic_evolutions}, because of reionization late-forming galaxies evolve very differently compared to early-forming galaxies and thus create a broadened stellar mass distribution. The bimodality most likely arises only because we assume that there is a sharp virial temperature cut-off which determines whether a certain galaxy can accrete baryonic matter. We checked this by analysing virial temperature dependence on $z$ for the model galaxies. Galaxies corresponding to the lower stellar mass component of bimodal distribution are strictly those for which virial temperature always remains lower than $T_\mathrm{eq}(\Delta_\mathrm{vir}/3)$. So we conclude that in reality this distribution should be only broadened but not bimodal because bimodality should be ``washed out'' by a smoother transition from non-accreting to accreting state. Because of this transition there is also a a sudden decrease in ratio between stellar mass and dark matter halo mass at $M_\mathrm{dm,0} \sim 10^{9 \mathrm{-} 10} \, M_\odot$. 
	
	In fig. \ref{compare_to_obs}, panel a, it is clear that at $M_{\rm dm,0} > 10^{11} \, M_\odot$, $M_{*,0}$ in our model grows slower with increasing $M_{\rm dm,0}$ than in both relations from abundance matching. This discrepancy might arise because in our model we assume that stellar feedback ejects gas from the galaxy completely so that it never falls back. In reality, at least some of ejected gas cools down and is eventually reincorporated back into the gas disk. So galaxies effectively accrete more gas and can form more stars. The importance of this effect increases with mass and so possibly explains the discrepancy in the higher mass end.
	
	In fig. \ref{compare_to_obs}, panel b, there is a group of outliers with up to 2 orders of magnitude higher $M_\mathrm{g,0}$ in the interval $M_{*,0} = 10^{5\mathrm{-}7} \, M_\odot$. These outliers are late-forming galaxies which experienced temporary quenching periods and accreted a significant fraction of their gas very late thus are extremely gas rich at $z=0$. It is important to note that the gas mass from \citet{cuspvscore, Kirby2017} includes only atomic hydrogen. In all of our models molecular gas fractions are very low (at most $\sim 1 \%$), but some fraction of disk gas might be ionized. This fraction should depend on stellar feedback and cooling rates but is not followed in our model and there is no simple way to estimate it. Therefore, the model gas mass shown should always be considered as an upper limit of atomic gas mass.
	
	In both panels c and d of fig. \ref{compare_to_obs} it is apparent that stellar and gas metallicity in models with $M_{*,0} > 10^7 \, M_\odot$ is too low compared to observed values. These discrepancies also arise because in our model the ejected gas never falls back. This leads to lower retained metal masses compared to those in real galaxies, and this discrepancy grows with increasing galaxy mass. In this work we are mainly concerned with galaxies having $M_{*,0} \leq 10^7 \, M_\odot$ (namely Leo A and Aquarius) where the agreement with observations is good even without incorporating the effects of gas fallback so we leave a detailed exploration of this effect for future works.
	
	The average gas metallicity decreases with increasing stellar mass at $M_{*,0} \sim 10^7 \, M_\odot$ (fig. \ref{compare_to_obs}, d) because a lot of lower mass galaxies are quenched by cosmic reionization and thus are left with the gas obtained at early times (if these are not evaporated already by $z=0$), which have been enriched during initial star formation. More massive galaxies, on the other hand, accrete more of the intergalactic gas at later times and thus their gas discs get diluted more. In our model infalling intergalactic gas always has primordial metallicity $Z_{\rm prim} = 2 \times 10^{-5}$, while in reality the average intergalactic metallicity should increase with time due to enrichment by gas ejected from galaxies. Inclusion of this effect should also decrease the metallicity discrepancy with observations at $M_{*,0} > 10^7 \, M_\odot$ as galaxies would accrete more metals throughout their history.
	
	It is also important to keep in mind that the gas metallicities in \citet{m_s_Z_g_relation} were determined by observing HII regions which correspond to sites of active star formation. Due to enrichment by forming stars in them these regions should have larger metallicities than the average metallicity of the disk. This also, at least partly, could explain the lower metallicity in our models because we only follow averaged metallicity throughout the whole disk. There is a similar problem when comparing stellar metallicities, as at least some of the stars form from gas enriched by massive stars that formed earlier in the same region. Also, \citet{m_s_Z_s_relation} used only red giant stars, which on average are older than the whole stellar population and so should represent chemical composition of a galaxy at earlier times. However, these effects should not produce very large changes and detailed modelling of them is beyond this work.
	
	\subsection{Comparison to Leo A and Aquarius galaxies}
	The star formation history of the blue model in fig. \ref{pic_evolutions}, e has similar features to the measured star formation histories of Leo A and Aquarius. In order to analyse whether these observed star formation histories can be explained by this model in more detail we calculated models for all isolated dark matter halos in our PINOCCHIO run with masses in the interval from $6 \times 10^9 \, M_\odot < M_\mathrm{dm,0} < 1.3 \times 10^{10} \, M_\odot$. This sample contains 1927 different galaxies. Outside of this interval similar star formation histories are extremely rare. To estimate the occurrence frequency of models similar to Leo A and Aquarius we counted how many of them form a fraction $f_{\rm late}$ of their stars only at $z < 1$ (later than $\approx 6$ Gyr after the Big Bang). We also counted only those galaxies which have $10^6 < M_{*,0} < 5 \times 10^6$ (similar to the values observed in Leo A and Aquarius, see table \ref{table_obs}). Out of all galaxies in this sample there are 16 ($0.8\%$) models with $f_{\rm late} \geq 0.7$ (similar to Leo A) and 53 ($2.8\%$) models with $f_{\rm late} \geq 0.5$ (similar or more extreme than Aquarius). We also checked and visually confirmed that the star formation histories of these models resemble the observed ones.
	
\begin{figure}
		 	\centering
		 	\includegraphics[width=9cm]{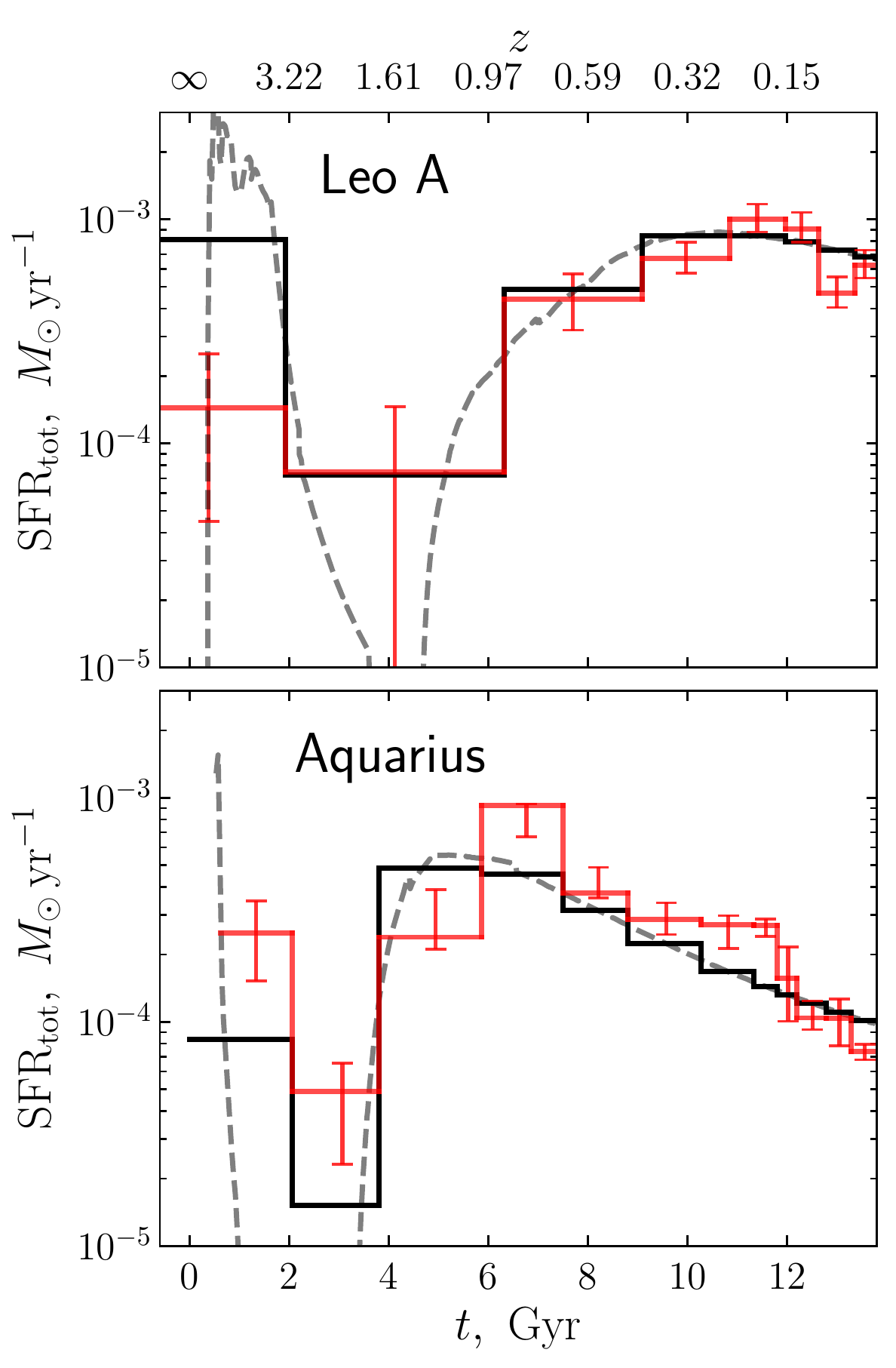}
		 	\caption{Example model star formation histories similar to those of Leo A (top) and Aquarius (bottom) dwarf galaxies. Grey dashed lines show star formation histories of the models, red lines show star formation histories determined from observations \citep{leo_sfh,aquarius_sfh} and black solid lines show star formation histories of the models averaged in same time intervals as the ones from observations for easier comparison.}
		 	\label{leo_aq_obs}
		 \end{figure}

    In fig. \ref{leo_aq_obs}, example models with similar star formation histories to those of Leo A and Aquarius are shown together with the ones derived from observations \citep{leo_sfh,aquarius_sfh}. The black solid lines show model star formation histories averaged in same time bins as ones from observations for easier comparison. The agreement between observations of Leo A and the model is very good: in all time bins except for the earliest one, the two agree to within 30\%, and the difference in the earliest bin is only a factor 5. For Aquarius the differences between the model and observations are larger, but in all time bins they do not exceed a factor $\sim 3$. So it is clear that the qualitative form of star formation histories in these models is very similar to the observed ones. Also almost certainly it would be possible to find better matching models if we would increase the sample of mass assembly histories. But we do not do that because we think that this qualitative agreement is enough to show that such star formation histories could be explained and it would be naive to expect that a semi-analytic model could reconstruct star formation histories of individual real galaxies in high detail. We stress that we do not need to fine-tune the parameters of the model in order to reproduce the qualitative features of the star formation histories of Leo A and Aquarius dwarfs. Rather, the set up of the model already provides such star formation histories as a consequence of including the effects of reionization on the baryon accretion into galaxies.

   The final dark matter halo, gas and stellar masses and average stellar metallicity of the same models as in fig. \ref{leo_aq_obs} are shown in table \ref{table_obs} together with the corresponding parameters of Leo A and Aquarius determined from observations \citep{Kirby2017}. The stellar masses of model galaxies are similar to observed ones, but gas mass in models is about 21 and 36 times higher than in Leo A and Aquarius respectively. We discuss the possible causes of this large discrepancy in section 4. The mass of metals in Aquarius is about 3 times higher than in model. This discrepancy probably is a result of the gas mass being too large, leading to dilution of enrichment.
	
	\begin{table*}
		\caption{Stellar masses, gas masses and average stellar metallicities of Leo A and Aquarius galaxies from \citet{Kirby2017} (metallicity is in [Fe/H]) and corresponding parameters at $z=0$ of the models that are shown in fig. \ref{leo_aq_obs}. Dark matter virial masses are only given for the models since to our knowledge they have not been determined observationally for these galaxies yet.}             
		\label{table_obs}      
		\centering          
		\begin{tabular}{c c c c c l l l }     
			\hline\hline\\[-1em] 
			\textbf{Galaxy}/model & $M_\mathrm{dm,0}$ & $M_{*,0}$ & $M_\mathrm{gas,0}$ & $\mathrm{[Fe/H]}_0$ \\ 
			& $(10^{9} \, M_\odot)$ & $(10^6 \, M_\odot)$ & $(10^6 \, M_\odot)$ &  \\ 
			\hline\\[-1em]                    
			\textbf{Leo A} & ... & $3.3 \pm 0.7$ & $7.4 \pm 0.8$ & $-1.67^{+0.09}_{-0.08}$ \\  
			Model & 8 & 5.2 & 160 & $-1.73$ \\
			\hline\\[-1em]
			\textbf{Aquarius} & ... & $1.5 \pm 0.2$ & $2.2 \pm 0.3$ & $-1.5 \pm 0.06$\\  
			Model & 7 & 2.3 & 79 & $-1.97$ \\
			\hline                      
		\end{tabular}
	\end{table*}

\section{Discussion}

	The results presented in the previous section show that for dark matter halos with present-day virial masses $M_\mathrm{dm,0} \sim 10^{9\mathrm{-}10} \, M_\odot$ the stochasticity from mass assembly histories is amplified by the reionization epoch in a sense that the late-forming halos are temporarily quenched while the early-forming halos are not quenched. This can result in qualitatively different star formation histories. Early-forming dark matter halos have the maximum star formation rate at $z \approx 4 \mathrm{-} 10$, after which time it monotonically declines. On the other hand, for most of the late-forming halos the star formation rate declines rapidly after the reionization epoch and then begins to rise again when they become massive enough for smooth baryon accretion to become again possible. In the latter scenario galaxies have minor old stellar populations and are dominated by stars which formed at $z < 2$. These features are qualitatively similar to those of some irregular dwarf galaxies, such as Leo A and Aquarius and fig. \ref{leo_aq_obs} shows that indeed this scenario could explain their star formation histories. However, as shown in table \ref{table_obs}, our model predicts that galaxies with these star formation histories should have $\sim 20 \mathrm{-} 30$ times more gas mass at $z=0$ than is observed in Leo A or Aquarius. This discrepancy is interesting because on average our models agree with observations quite well (fig. \ref{compare_to_obs}), at least for galaxies with $M_* < 10^7 \, M_\odot$. This could mean that in galaxies with delayed star formation histories like Leo A and Aquarius there are some additional, maybe external, processes that are not accounted for in our model. One possible solution would be inclusion of modelling the ionized gas fraction in the galaxy, because now we compare total gas mass in our models with only the observationally-constrained atomic gas mass. Other more exotic solution would be to include the effects of active galactic nuclei (AGNs). Recently, evidence for active intermediate mass black holes has been found in ever smaller dwarf galaxies \citep{AGN_in_dwarfs_detections} and there are no known reasons why it would be impossible for them to exist in even smaller galaxies like Leo A or Aquarius. This is of course speculative, but worth considering as feedback from AGNs could at least partially solve a lot of known problems with dwarf galaxies \citep{Silk_AGN_in_dwarfs}. In our case, AGN winds or jets could increase the star formation rate by generating turbulence and compressing gas \citep{AGN_enhanced_SFR_obs, AGN_enhanced_SFR_mod} and also could easily eject significant amounts of gas out of low-mass galaxies like Leo A and Aquarius. Interestingly, both of these galaxies show large holes in their HI maps\footnote{https://science.nrao.edu/science/surveys/littlethings/data} \citep{LITTLE_THINGS}. Such gas ejection might result in galaxies with similar star formation histories but with significantly lower gas masses at $z = 0$, thus solving the discrepancy. We plan to investigate this possibility in the future. 
	
	Other formation scenarios, where these ``late-blooming'' galaxies form early and then do not form stars for a very long time because of absence of any perturbations or where they assemble their mass extremely late seem less likely because such growth histories are extremely rare according to structure formation simulations based on $\Lambda$CDM cosmology (we do not find any examples of these in our whole sample of merger trees). So it seems that the effect of reionization as described in this work is a more natural explanation. Of course, even though we find models with similar star formation histories to Leo A and Aquarius, the fraction of these models is quite small (see section 3.3) and these models are clearly outliers. However, it might be that Leo A and Aquarius are also outliers in reality. So to fully answer the question, whether our model can successfully explain their evolution, a statistical abundance investigation and comparison is needed. The gas mass discrepancy that we find could also be thought of as evidence that the standard $\Lambda$CDM cosmology fails at small scales \citep[for a review of more evidence for this see][]{small_scale_problems_LCDM}. But this claim is very strong and more conventional solutions should be sought at first. 
	
	Other works have also found that dwarf galaxies with strongly delayed star formation histories tend to reside in late-forming $10^{9\mathrm{-}10}$ halos \citep{dopey_and_grumpy, reborn_dwarfs_1, reborn_dwarfs_2}. In all these works it was also found that reionization probably plays a very important role in explaining how dwarf galaxies with two component stellar populations (old and young) could be formed. Another work, published while our paper was in review, shows that similar reignition of star formation in dwarf galaxies could occur when they collide with cosmic web or tidal gas streams left over from galaxy mergers \citep{reignition_of_dwarfs_tidal_tails}. These galaxies also form in late-forming dark matter halos with similar masses as in this work and also tend to have significantly higher gas masses as opposed to other similar mass dwarfs. So it seems that there might be several mechanisms at play in the formation of galaxies like Leo A and Aquarius and because of similar resulting observational properties they could be hard to separate.
	
    We also find that some $M_\mathrm{dm,0} \lesssim 10^{10} M_\odot$ models are quenched for so long that their star formation rate begins to increase only at $z \sim 0$ (see cyan model in fig. \ref{pic_evolutions}). In these cases galaxies end up with stellar masses $M_{*,0} \sim 10^{5 \mathrm{-} 6} \, M_\odot$ and their baryonic mass is completely dominated by the gas with $M_{\rm gas,0} \sim 10^{7\mathrm{-} 8} \, M_\odot$. This might explain the existence of some observed extremely gas-dominated dwarf galaxies \citep[e.g., ][]{gas_dominated_dwarf}. In addition, our model predicts a steepening decrease in number of galaxies with $M_{*,0} < 10^6 \, M_\odot$ (fig. \ref{compare_to_obs}, a) because of the effect of reionization on low-mass halos. This might help solve the missing satellites problem.
	
	Of course the presented model is highly simplified. Like in other semi-analytic galaxy formation models, a major weakness is the modelling of stellar feedback as it involves a complex interplay between many different processes on different scales. However in our knowledge presently there are no more realistic and practical ways to model it analytically and we must wait for progress in this area. Another major simplification is that we do not take into account the instabilities in the gas disc which can occur during major mergers. In the model during mergers with non-empty dark matter halos the star formation rate increases only because of rapid increase in the gas mass. In reality these merger-induced instabilities could lead to starbursts which might even drive the gas out of the dark matter halo and leave the remaining galaxy quenched. As mentioned in section 2.3, to study how these starbursts could affect our results we calculated models in which extreme starbursts happen during major mergers (mergers with mass ratios of 30\% or higher). We assume that during these starbursts a fraction of the gas is turned to stars and the remaining gas is ejected, proportionally to formed stellar mass so that in total all of the gas would be used up. Certainly this treatment is too extreme but it is well suited to study how our results would change by including more realistic major merger physics as the changes then should only be smaller than with this extreme adjustment. We find that including extreme starbursts does not change the final parameter distributions strongly. In most of the models major mergers happen very early (at $z \gtrsim$ 2) so the effect is washed out by later evolution. Of course in some models with late major mergers the star formation is quenched and so their star formation history is altered strongly but these models constitute only minority. In addition we find that majority of temporary quenched models preserve their star formation histories similar to Leo A and Aquarius. From these results we conclude that inclusion of merger-induced starbursts should not change our main results and conclusions. 
	
	We also calculated models with different gas disk scale length compared to stellar disk (section 2.3), namely with $R_\mathrm{s,g} = R_\mathrm{s,*}$, $1.5 R_\mathrm{s,*}$ and $3 R_\mathrm{s,*}$. Lower scale lengths lead to more concentrated gas disks and thus higher star formation rates. On average this does not affect the final parameter distributions significantly so letting it to vary would only increase their dispersion. It does however change the star formation histories of individual galaxies. With lower scale lengths galaxies tend to form their stars earlier and with larger scale lengths later. So by varying this scale length ratio it would be possible to find better-matching models to Leo A and Aquarius. We also find that models with $R_\mathrm{s,g} = R_\mathrm{s,*}$ that have similar star formation histories to Leo A and Aquarius have $\sim 10$ - 20 times more gas mass than is observed thus reducing the discrepancy by a factor of about $2$ but still cannot solve it completely. It is important to keep in mind that to our knowledge there is no reason that the ratio between scale lengths of gas and stellar disks should be constant throughout the whole evolution of a galaxy. That might also have important effects to resulting star formation histories but we leave this question for future research. 
	
	Nevertheless with all these shortcomings the model still predicts galaxy parameters similar to the observed ones (fig. \ref{compare_to_obs}). And it is important to note that this is achieved, contrary to other semi-analytic galaxy formation models, without optimizing any free parameters. All the coefficients and parameters (e.g. star formation efficiency, clumping parameter, momentum generated by supernovae) are taken from observations or numerical simulations and are held fixed. The agreement between this model and observations could be easily improved by finding optimal values for these parameters. But we choose not to do it because it would decrease the predictive power of the model. Our simple model works because in this article we study probably the most simple type of galaxies: isolated and low-mass. As galaxy formation and evolution is a very complex process incorporating many different processes on very different space and time scales, it might be very useful to first understand the most simple systems and only then go to more complex large galaxies. 

\section{Conclusions}

	In this work we used a semi-analytic model to study the effect of mass assembly stochasticity on the evolution of isolated dwarf galaxies with present day virial masses $M_\mathrm{dm,0} \sim 10^{9 \mathrm{-} 11} \, M_\odot$. The main conclusions of this work are the following:
	
	1) The interval of dark matter halo virial masses at $z = 0$ that is studied in this work ($10^{9\mathrm{-}11} \, M_\odot$) includes the characteristic transitional mass below which halos are strongly affected by cosmic reionization and above which they are affected weakly. 
	
	2) At this mass, because of the different possible mass assembly histories, a fraction of galaxies are strongly affected by reionization while the remaining ones are affected weakly. This can result in qualitatively differing star formation histories of different galaxies with the same dynamical mass at $z = 0$.
	
	3) The models in which dark matter halos assemble their mass later than with average mass assembly history have at least qualitatively similar star formation histories to Leo A and Aquarius dwarf galaxies.
	
	4) However these models tend to have 20 - 30 times more gas mass than is observed, thus indicating that some additional processes should be included to completely reconstruct these galaxies.

	The results also show that isolated dwarf galaxies can be modelled quite successfully by using semi-analytic models without optimizing any free parameters. This might be very useful in studying large parameter volumes as semi-analytic models require orders of magnitude less computational resources than 3-D hydrodynamical models. The presented model has some major approximations but it is possible to relax or soften at least some of them and make the model even more realistic in the future.

\begin{acknowledgements} We thank the anonymous referee for helpful comments and thought provoking questions. We also thank V. Vansevičius, D. Narbutis and P. Monaco, the creator of PINOCCHIO, for helpful discussions and comments. This research was funded by a grant (No. LAT-09/2016) from the Research Council of Lithuania. \end{acknowledgements}

\bibliographystyle{aa.bst} 
\bibliography{references.bib} 

\end{document}